\documentclass[journal=jacsat,manuscript=article]{achemso}

\usepackage[utf8]{inputenc}
\usepackage[brazil,english]{babel}
\usepackage{csquotes}
\usepackage{lineno,hyperref}
\usepackage{amsfonts}
\usepackage{amssymb}
\usepackage{float}

\usepackage{chemformula} 
\usepackage[T1]{fontenc} 
\usepackage{graphicx}
\usepackage{epsfig}
\usepackage{subfigure}
\usepackage{setspace}
\usepackage{amsmath}
\usepackage{color}
\usepackage{gensymb}
\usepackage{booktabs}
\usepackage{multirow}
\usepackage[normalem]{ulem}
\usepackage{soul}
\author{W. H. S. Brandão}
\affiliation{Departmento de F\'isica, Universidade Federal do Piauí, Teresina, Piauí, Brazil.}
\email{wjefferson.henrique@ufpi.edu.br}

\author{A. L. Aguiar}
\affiliation{Departamento de F\'isica, Universidade Federal do Piau\'i, 64049-550 Teresina, Piau\'i, Brazil}
\email{acrisiolins@ufpi.edu.br}

\author{J. M. De Sousa}
\affiliation{Instituto Federal do Piauí - IFPI, São Raimundo Nonato, Piauí 64770-000, Brazil.}
\email{josemoreiradesousa@ifpi.edu.br}

\title{Atomistic Computational Modeling of Temperature Effects in Fracture Toughness and Degradation of Penta-graphene Monolayer}

\begin{document}

\section{Abstract}
The novel carbon allotrope with particular and unique
2D arrangement of carbon atoms similar to
a Cairo pentagonal tiling, with interplay of $sp^{3}$ and $sp^{2}$ hybridized carbon atoms is called of Penta-graphene (PG). Previous theoretical investigations have shown that PG monolayer is mechanically and thermodynamically stable, possessing also a large band gap of $3.25eV$.
This new carbon allotrope with unique carbon atom arrangement in a network (non-coplanar pentagons) is the focus of the theoretical investigations in this work.
Using the non-equilibrium molecular dynamics simulations with reactive modern force field ReaxFF, we performed computational modeling of the nanostructural, dynamics e mechanical properties of penta-graphene monolayer under high temperature conditions.
We obtained in our results the effect of the temperature in mechanical properties of penta-graphene monolayer up to $2000K$, where our results show that strain rate was strong effect on the mechanical properties with reduction of the $67$\%, reduction  in the Ultimate Tensile Streght (UTS) $ 35.88 - 11.83GPa.nm $ and Young's Modulus ($Y_{Mod}$) of the $ 227.15 - 154.76GPa.nm $. 

In this work we also calculated the reactive degradation of monolayer of penta-graphene at temperatures changes of $10K$ up to $2000K$.
Thus, our averages show that penta-graphene monolayer loss atomic configurations with temperature effect up to $600K$, where the monolayer show nanostructural transition with several islands of graphene, large regions of porosity, small 1D carbon chains, and also negative curved layer.

\section{keyword}
Penta-graphene Monolayer, Mechanical Properties, Temperature Effects, Reactive Molecular Dynamics Method, Nanotechnology, Fracture

\section{Introduction}
It is of great interest of nanotechnology, as as an important and fundamental sub-area of material science, the understanding of the physical/chemical properties of nanomaterials and how they behave under different environment conditions at nanometric scale. \cite{mazzola2003commercializing,bhattacharya2005nanotechnology,fedorov2020nanotechnology}.
Since the graphene extraction\cite{novoselov2005two}, a lot of theoretical and experimental works brought us closer to improving quality in applicability and development of new materials with superior qualities to the existing ones \cite{zhou2012oxygen,jastrzkebska2012recent,jan2014improvement}. Transistors in the improvement of electron-device \cite{schwierz2010graphene}, graphene with applications for conventional plasmonics fields \cite{grigorenko2012graphene}, graphene-based composites in $Li$-ion batteries applications, fuell cells, photovoltaic devices, supercapacitors and photocaralysis graphene-based materials \cite{huang2012graphene,li2008graphene}, are examples that, without an electronic gap, graphene-based semiconductor nanodevices have very limited applicability. \cite{withers2010electron}.

\begin{figure}[htb!]
 \centering
 \includegraphics[scale=0.35]{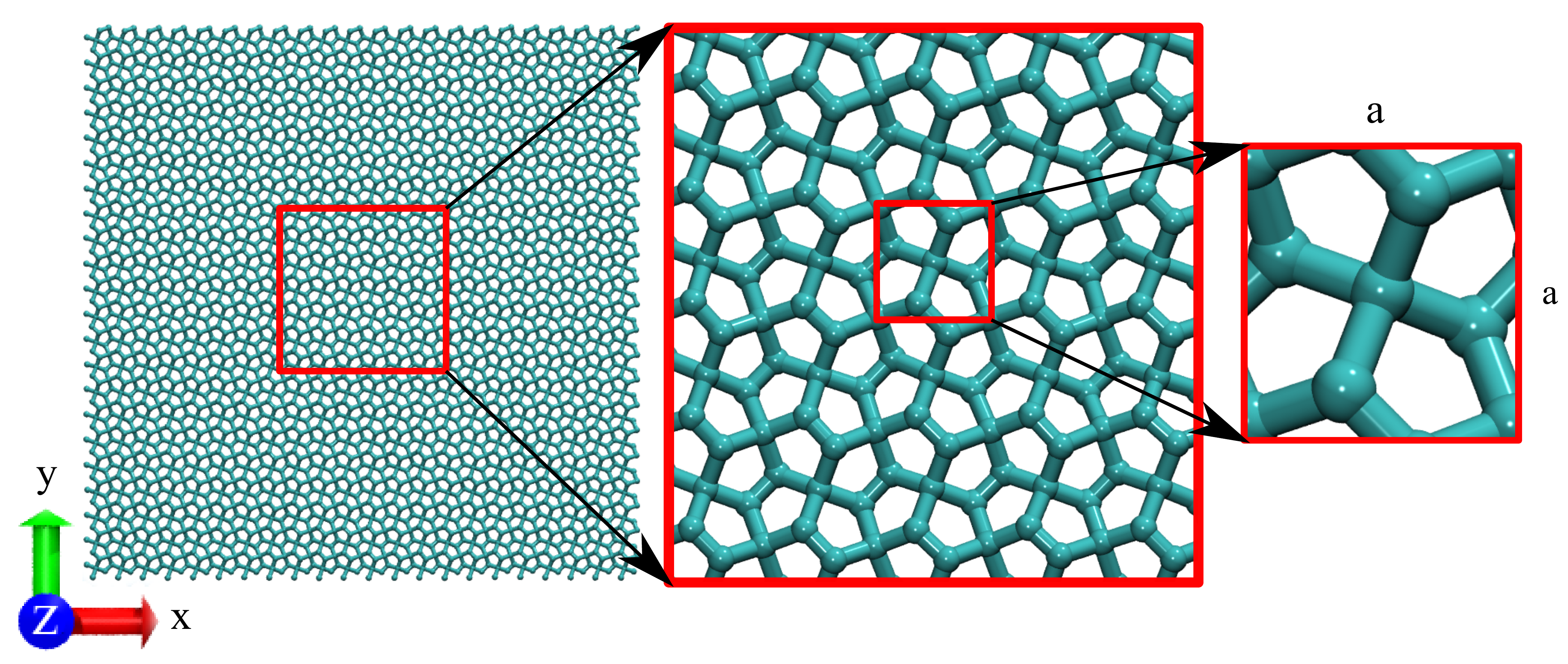}
 \caption{Atomistic model of Penta-graphene (PG) monolayer, with 
square periodic boundary conditions ($80$ \AA $\times$ $80$ \AA). Middle and right panels represents respectively a zoomed view of PG monolayer and its unit cell ($a = 3.64$~\AA) composed by $5$ carbon atoms.}
\label{fig:PentaG}
\end{figure}

Nevertheless, new exotic carbon allotropes have been intensively investigated, with nanoelectronic characteristics equal or superior to graphene. Such as, graphynes, nanostructures planar forms of carbon with phases containing $sp^{2}$ and $sp$ carbon atom hybridization \cite{baughman1987structure}, $T$-carbon, a stable crystalline carbon allotrope whith lower density $1.50 g/cm^{3}$ and direct band gap about $3.0 eV$ \cite{sheng2011t}, D-carbon, a crystalline orthorhombic  carbon allotrope whith hybridization $sp^{3}$ in the space group $Pmma$ $[D2h5]$ with $6$ carbon atoms in the unit cell, with a band gap of $4.33 eV$ \cite{fan2018d}. We also can cite the Orthorhombic $C14$, a novel carbon allotrope with $sp^{3}$-bonded networks with a $14$ carbon atom in the orthorhombic unit cell and direct band gap of $4.60 eV$, twin graphene, a carbon allotrope (in the space group of $P6/mmm$) with $18$ carbon atoms in unit cell and electronic band gap of about $1.0 eV$ \cite{jiang2017twin}. Other important examples is the $\psi$-graphene, a metallic carbon allotrope composed of $5-6-7$ carbon rings for applications as anode in batteries \cite{li2017psi}, the popgraphene, a novel carbon allotrope with $5-8-5$ carbon atoms rings where is intrinsically metallic \cite{wang2018popgraphene}, the carbon allotrope called Hp-$C17$, a nanostructure with all-$sp^{3}$ network formation, where the unit cell are composed of $17$ carbon atom of space group of $P6m2$, with indirect band gap of $4.04 eV$ \cite{su2020hp}, and finally a novel 2D carbon allotrope called Penta-graphene (PG), formed by pentagons arrangement of carbon atoms similar to
a Cairo pentagonal tiling, mechanically (with negative Poisson's ratio) and thermodynamically stable with large band gap of $3.25$eV  \cite{zhang2015penta}. 
Atomistic model of Penta-graphene monolayer can see in Fig.\ref{fig:PentaG}.

In Fig.\ref{fig:PentaG} we show an atomic model of monolayer of penta-graphene with bidimensional configuration of $80 \times 80$ \AA, where a small box show the unit cell ($a = 3.64$~\AA). Previous theoretical investigations show that penta-graphene monolayer has a versatility of the applications in nanoelectronics and nanomechanics. {\it Wu, X. et. al.} ($2016$) show that penta-graphene monolayer hydrogenated increase the thermal conductivity of ($615 W/mK$)  than that of penta-graphene monolayer ($350 W/mK$) \cite{wu2016hydrogenation}. {\it Cranford, S.} ($2016$), show that penta-graphene monolayer has stiffness (approximately $378.3 N/m$),  thermal energy transition for temperature up to $600K$  \cite{cranford6to5}. { \it Sun, H. et. al.} ($2016$) show that mechanical  behavior  of  monolayer  penta-graphene using DFT calculations have lower Ultimate Tensile Streght and Young's Modulus compared to graphene monolayer \cite{sun2016mechanical}. {\it Li, X. et. al.} ($2016$) show that penta-graphene monolayer hydrogenated and fluorinated penta-graphene sheets can effectively tune the electronic and mechanical properties  \cite{li2016tuning}. {\it Yuan, P. F. et. al.} ($2017$) show that penta-graphene bidimensional have physical characteristic of bipolar  magnetic  semiconducting features \cite{yuan2017electronic}. {\it De Sousa, J. M. et. al.} ($2020$) show that Penta-graphene monolayer at room temperature have about $20$\% of strain whith nanofracture dynamical bond breaking formation of $7$, $8$ and $11$ carbon atoms rings and chains, Young's Modulus of about $206GPa.nm$, Ultimate Tensile Streght about $32GPa.nm$ \cite{de2017mechanicalPentaGraphene}. Studies show that penta-graphene monolayer have Young's modulus in the range of 133.9 - 322.0 GP.nm with MD calculations (reaxFF at room temperature) and 257.6 via DFT\cite{desousa2021penta}. Nanotubes based in penta-graphene geometry have been studied as well. {\it Chen, M. et. al.} ($2017$) show that penta-graphene nanotubes have plastic  characteristics with strain-rate and nanotube-diameter independence \cite{chen2017mechanical}. {\it Quijano-Briones, J. J. et. al. } ($2017$) summarize results about phonons dispersion frequencies calculations  penta-graphene nanotubes \cite{quijano2017chiral}. {\it De Sousa, J. M. et. al.} ($2018$) studied zigzag-like and armchair-like penta-graphene based nanotubes of different diameters, where Young's Modulus is about $800 GPa$ with distinct elastic behavior in relation to conventional carbon nanotubes \cite{de2018mechanical}.

However, a more detailed study of the physical properties of penta-graphene monolayers is still needed. Thus, the purpose of this work is to show that penta-graphene membranes are sensitive in mechanical properties when we change  temperature, where critical strain $\sigma_{critical}$ decrease as function fo temperature $300K$ - $20$\%, $600K$ - $13$\% and $900K$ - $6$\%, for example. The Young's Modulus change $ 227.15 - 154.76GPa.nm $ and Ultimate Tensile Streght $ 35.88 - 11.83GPa.nm $. Our results show also that penta-graphene monolayer have transitions configurations at effects temperatures between $10K$ up to $2000K$, where our observation show islands of graphene, distint arrangement atomic configurations, chains and negative curved monolayer.

\section{Computational Methodology}

All reactive molecular dynamics simulations has carry out performed by non-equilibrium molecular dynamics method \cite{alder1959studies,allen1989computer,rapaport2004art}. The numerical atomic positions os carbon atoms are calculated by large-scale atomic/molecular massively parallel simulator (LAMMPS) code \cite{plimpton1995fast}. The interatomic potential used in all calculations are a modern reactive force field ReaxFF with set parameter described by references \cite{van2001reaxff,mueller2010development}. The reactive force field ReaxFF is parameterized using available experimental results and first-principles calculations \cite{van2001reaxff,mueller2010development,chenoweth2008reaxff,raymand2008reactive,jarvi2008development, liu2011reaxff,senftle2016reaxff}. Approximately to empirical non reactive force fields, ReaxFF is divide by partial energy contributions, as in the follow equations \ref{eq:energy-reax} \cite{van2001reaxff}:
\begin{eqnarray}
\label{esys}
E_{system}&=&E_{bond}+E_{over}+E_{under}+E_{val}\nonumber \\
&&+E_{pen}+E_{tor}+E_{conj}+E_{vdW}\nonumber\\
&&+E_{co} \quad , 
\label{eq:energy-reax}
\end{eqnarray}
where, here the terms of Eq.\ref{eq:energy-reax}, respectively, represents the energies corresponding to the bond distance $(E_{bond})$, the over-coordination $(E_{over})$, 
the under-coordination $(E_{under})$, the valence $(E_{val})$, the penalty for handling atoms with two double bonds $(E_{pen})$, the torsion $(E_{tor})$, the conjugated bond energies $(E_{conj})$,  the van der Waals $(E_{vdW})$, and coulomb interactions $(E_{co})$. A  theoretical fundametation of ReaxFF is defined by bond order $BO'ij$ between a pair of atoms as show in Eq.\ref{eq:bond-order} follow \cite{van2001reaxff}:\\
\begin{eqnarray}
BO'ij = exp\left[ p_{bo,1} \cdot \left( \frac{r_{ij}}{r_{o}} \right)^{p_{bo,2}} \right] + exp\left[ p_{bo,3} \cdot \left( \frac{r_{ij}^{\pi}}{r_{o}} \right)^{p_{bo,4}} \right] + \nonumber\\ 
+ exp\left[ p_{bo,5} \cdot \left( \frac{r_{ij}^{\pi \pi}}{r_{o}} \right)^{p_{bo,6}} \right] 
\quad, 
\label{eq:bond-order}
\end{eqnarray}
where the atomic configurations is obtained from interatomic distance $r_{ij}$ of three exponential terms, such as, the $\sigma$ bond $(p_{bo,1})$ and $(p_{bo,2})$, first $\pi$ bond $(p_{bo,3})$ and $(p_{bo,4})$ and $\pi\pi$ bond $(p_{bo,5})$ and $(p_{bo,6})$, with their respective dependencies in interatomic distances $C - C$ bond ($ \sigma  \sim 1.5 $\AA), ($\pi \sim 1.2 $\AA) and  ($\pi\pi \sim 1.0 $\AA). A lot of theoretical reactive molecular dynamics simulations has been performed by CMD-ReaxFF \cite{bagri2010structural,burtch2014water,autreto2014site,de2016torsional}.

In this work, we study the nanostructural and nanofracture dynamics  of penta-graphene monolayer under temperature effects. Therefore, we considered in this work square membranes under periodic boundary conditions with dimensions of approximately 
$80$\AA$\times80$\AA$~$ whith $2904$ carbon atoms. We initially, to eliminate the residual stress on the penta-graphene monolayer, we carried out an energy minimization followed by a isothermal-isobaric $(NPT)$ integration on Nose-Hoover style non-Hamiltonian equations of motion of carbon atoms turn null press in monolayer of penta-graphene, before starting the penta-graphene monolayer stretching processes. The load mechanical strength calculations were performed by canonical $(NVT)$ ensemble at $10K$ up to $2000$, stretching the penta-graphene monolayer until nanofracture characterized by the complete rupture of the membrane in two parts. The updates of carbon atoms position and velocity in the bidimensional nanostructure studied in this work in each timestep of simulations are described by Nosé-Hoover Thermostat \cite{nose1984molecular,nose1984unified,evans1985nose,morishita2010nose}. Thus, through this thermostat, the penta-graphene membranes were heated from low to high temperatures, ranging from $10K$ up to $2000K$. 

The constant engineering tensile strain rate used for stretching velocity of penta-graphene monolayer are $\delta = 10^{-6}/fs$, where the dimensions of length uniaxial strain of penta-graphene monolayer are change $L(t)=L_{0}(1+\delta t)$: ($L_{0}=80$\AA - length of  penta-graphene monolayer streght in $x$ direction). This constant engineering tensile strain rate is small enough to characterize and analyze in detail the nanofracture nanostructural in the stretching dynamics and degradation of penta-graphene monolayer.  In all molecular dynamics simulations performed in this work, the calculations were obtained, with timestep of numerical integration of $0.05$ fs. The stretching process dynamics was calculated by virial stress tensor along the stretching direction $x$, described by the following equation \ref{eq:stress-virial}:

\begin{eqnarray}
  \sigma_{\alpha \beta }&=&\frac{1}{\Gamma }\sum_{i}^{N}\left (m_{i}v_{\alpha i}v_{i\beta } + r_{i\alpha }f_{i\beta } \right ) = \nonumber\\
 \sigma_{x} &=& \sigma_{xx }=\frac{1}{\Gamma }\sum_{i}^{N}\left (m_{i}v_{ x}v_{i x } + r_{ix }f_{ix} \right), 
  \label{eq:stress-virial}
\end{eqnarray}
where $\Gamma$ is the volume of penta-graphene monolayer, $N$ the number of carbon atoms, $m_{i}$ the mass of carbon atom, $v$- velocity, $r$- spatial position of carbon atoms, $f$- the force per carbon atom. The elastic properties of the penta-graphene monolayer were analyzed by the stress-strain relationship described by the following equation \ref{eq:young-mod}, 
\begin{eqnarray}
\label{eq:young-mod}
\centering
\varepsilon &=& \frac{\zeta - \zeta _{0}}{\zeta _{0}}= \frac{\Delta \zeta }{\zeta _{0}} =  \nonumber\\
Y_{Mod} &=& \frac{\sigma_{x}}{\varepsilon _{x}},
\end{eqnarray}
where $\zeta_0$ and $\zeta$ are the length of penta-graphene monolayer before and after the dynamics of stretching, respectively. So, the Young's modulus is obtained by the stress/strain ratio in the elastic regime of the monolayer of penta-graphene in $x$ direction (Eq.\ref{eq:young-mod}). In order to perform a more detailed analysis of the distribution of stress along the structure during the fracture process, we also calculated the quantity known as {\it von Mises stress}, $\sigma_{vM}$, which is mathematically given by follow equation Eq.\ref{eq:stress-tensor}~\cite{garcia2010bioinspired}:
{\small
\begin{eqnarray}
  \sigma_{vM}=\left [ \frac{(\sigma_{xx}-\sigma_{yy})^{2}+
      (\sigma_{yy}-\sigma_{zz})^{2}+(\sigma_{zz}-\sigma_{xx})^{2}+
      6(\sigma_{xy}+\sigma_{yz}+\sigma_{zx})^{2}}{2} \right ]^{\frac{1}{2}} , 
      \label{eq:stress-tensor}
\end{eqnarray}
}
where $\sigma_{xy}$, $\sigma_{yz}$ and $\sigma_{zx}$ are shear stress components. 
This methodology is widely used in atomistic simulations involving nanostructured systems formed by atoms of carbon~\cite{wang2007mechanical,dos2012unzipping,coluci2007atomistic,bizao2017mechanical}, as well as hybrid C and N systems~\cite{de2016mechanicalCN} and, silicene~\cite{botari2014mechanical}, for instance. It allows a systematic way to visualize how the stress accumulates and dissipates during the stretching/fracture processes along the whole of penta-graphene monolayer. In all molecular dynamics simulations performed in this work we use the VMD \cite{humphrey1996vmd} for the visualization of the Von Mises tensor stress fields.

\section{Results and Discussion}

\subsection{Thermal degradation}

\begin{figure}[htb!]
 \centering
 \includegraphics[scale=0.30]{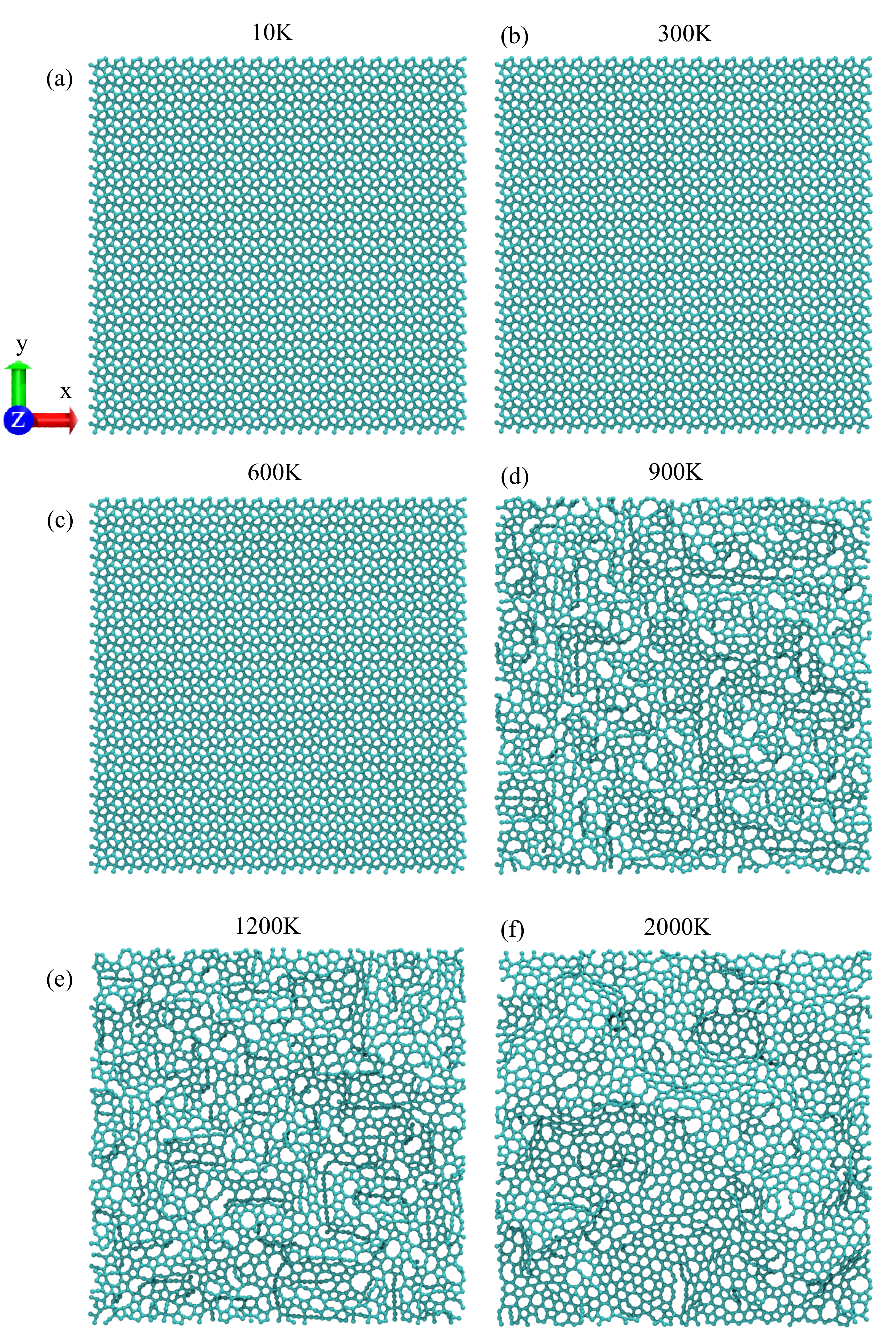}
 \caption{Atomistic representative frames of reactive molecular dynamics simulations of heating degradation of penta-graphene monolayer. The temperatures are (a) $10K$, (b) $300K$, (c) $600K$, (d) $900K$, (e) $1200K$ and (f) $2000K$. The total time of penta-graphene heating was $1250$ picoseconds.}
\label{fig:effects-temp}
\end{figure}

Firstly, Fig.\ref{fig:effects-temp} shows a sequence of the temporal frames from computational modelling of penta-graphene monolayer heating from $10K$ up to $2000K$. Our results show that for temperatures below $600K$ (see Fig.\ref{fig:effects-temp} (a), (b) and (c)) the penta-graphene monolayer has its atomic configuration with carbon atoms arranged in a network (non-coplanar pentagons) preserved. However, at $900K$, the penta-graphene monolayer loses its original symmetric configuration as one case see in Fig.\ref{fig:effects-temp}d. At this point, we can see several larger pores and non-planar curved regions, but despite that observed porosity, we can still see some original non-coplanar pentagons islands preserved. In Fig.\ref{fig:effects-temp} (e), at 1200K, one can observe the increasing of porosity in the penta-graphene monolayer. 

\begin{figure}[htb!]
 \centering
 \includegraphics[scale=0.38]{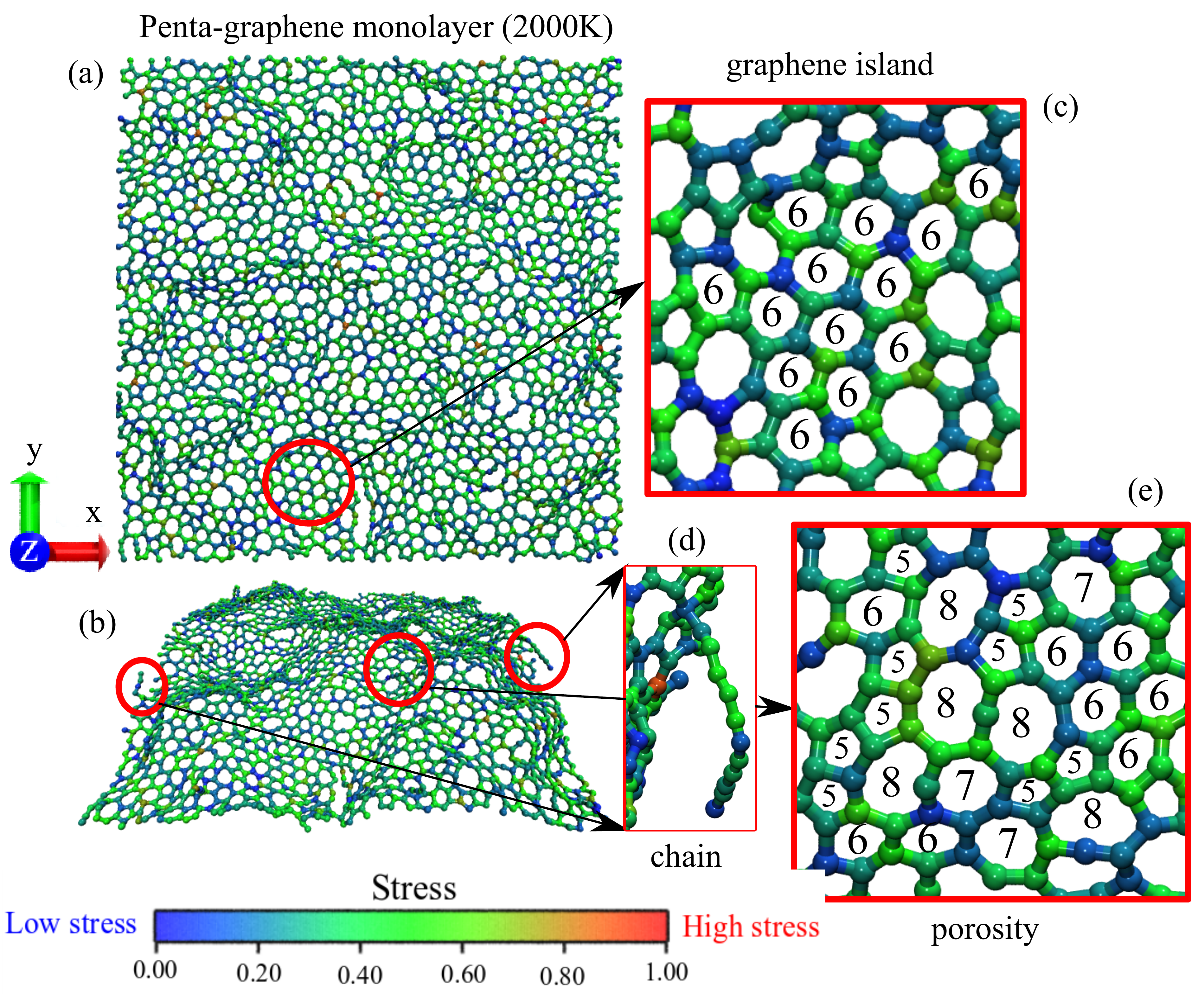}
 \caption{Atomistic representative frames snapshots of effects temperature in degradation of penta-graphene monolayer at $2000K$. In (a) the penta-graphene monolayer thermalized at 2000K. In (b) a perspective view of the penta-graphene monolayer. In (c) a zoomed show graphene islands. In (d) some chains and (e) porous formations. The color bar represents the level of von mises stress, where the color blue (red) represents low (high) stress, respectively. The time of heating of the penta-graphene monolayers was $1250$ picoseconds.}
\label{fig:effects-2000K}
\end{figure}

At 2000K, (see Fig.\ref{fig:effects-temp} (f)), it is interesting to see that there is atomic rearrangement where it emerges several graphene islands and, as well as porous, and some small carbon chains. Details of graphene
clusters formed at 2000K and also octagons, heptagon porous can be visualized in Fig.\ref{fig:effects-2000K}. Longer carbon chains are also observed
over the membrane surface. Therefore, we observed that at higher temperatures (near 2000K) carbon atoms have enough kinetic energy to move easily over the membrane surface, reducing residual strained conformation of curved penta-graphene islands, and the membrane converges approximately to a more stable atomic arrangement, close to graphene and other pure planar carbon arrangements. Cranford et al.\cite{cranford6to5} have observed that potential energy of penta-graphene increases nominally until 600 K (without any reconstruction), then the systems start the structural transition to graphene, which saturates approximately at 2000 K, which are in agreement with our findings.

\begin{figure}[htb!]
    \centering
\includegraphics[scale=0.60]{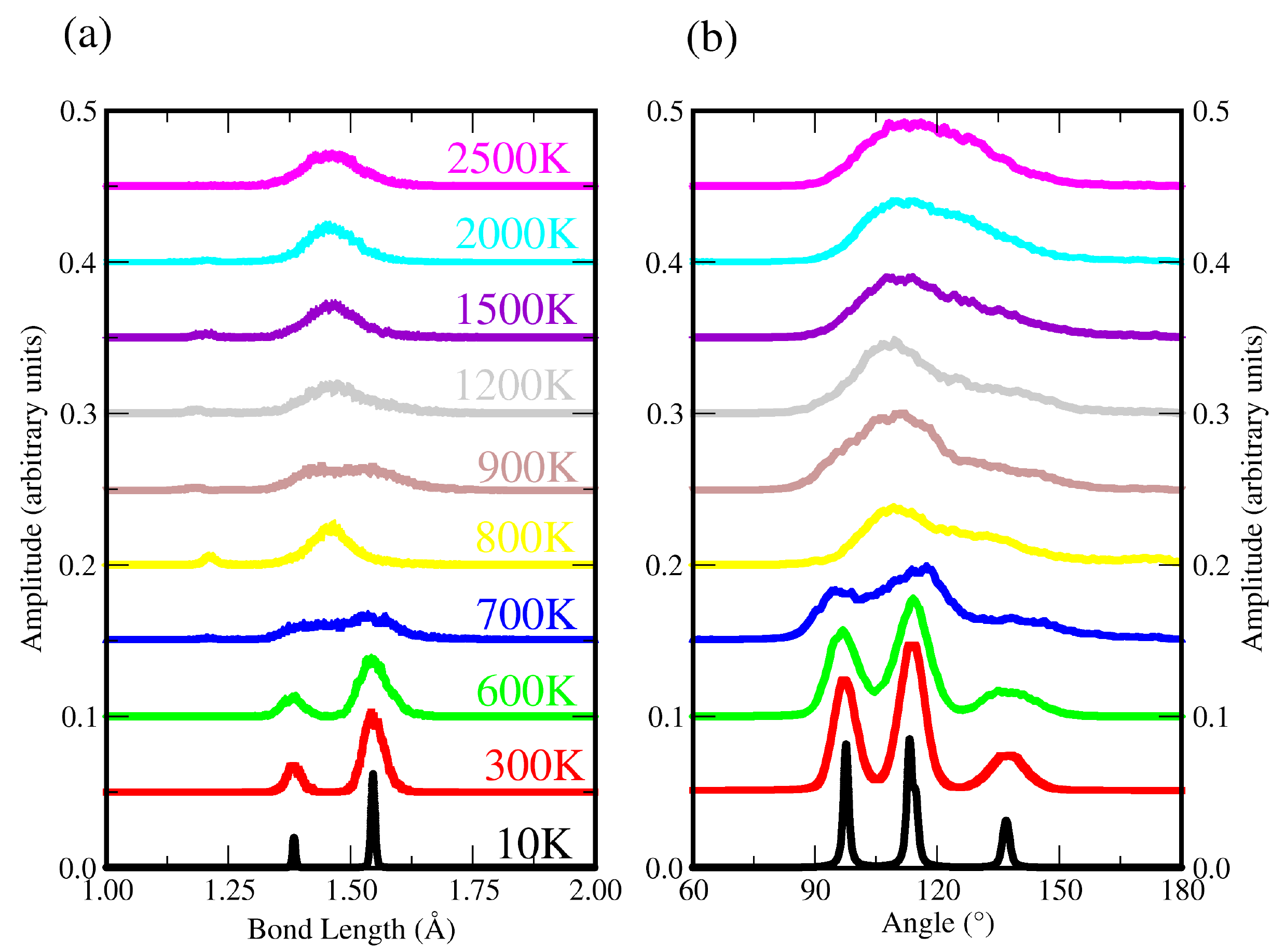}
    \caption{Bond (a) and angle (b) distribution function for PG membranes on simulated temperatures. The characteristic low temperature peaks disappear for temperatures above 600K (green curve).}
    \label{fig:angle-bond-dist}
\end{figure}

Details of structural evolution of the PG membranes as influenced
by temperature, can be see in Fig.\ref{fig:angle-bond-dist} as the analysis of C-C bonds and angle distributions collected at final stage
of thermal bath. Fig.\ref{fig:angle-bond-dist}(a) shows the bond distribution at 10K, where we can observe two narrow peaks due two non-symmetric characteristic bonds of pentagraphene (1.38 and 1.55 \AA). For temperatures 300K 
and 600K we observed the broadening of those two bonds, which
is expected from natural bond vibrations. These peaks are no longer clearly distinguishable in the transition between 600 to 900 K with the temperature degradation of the membrane. After 1200K, a for temperatures higher, we clearly observed a broad peak centered at around 1.40 \AA which is characteristic
of sp2-like carbon materials. Similarly for the distribution of angles (\figurename{ \ref{fig:angle-bond-dist}(b)}), there are three characteristic angles peaks in the structure (identified at 97.70, 113.36 and 137.01 \AA) at 10K. The intensity of these peaks decreases, and become larger in
width with the increasing of temperature up to 600-700K, 
where completely spread after 900 K. 

\subsection{Stress-strain relations at high temperatures}

In order to investigate thermal effects on elastic properties of PG membrane,
we have study the mechanical behavior of the monolayer under strong stretching
conditions up to the limit of rupture. In Fig.\ref{fig:ss-all-temp} we show the stress-strain curves obtained from classical molecular dynamics simulations with increasing temperature. Since we have showed in previous section that PG membrane
lose its original structural symmetry between 600K and 900K, we investigated the thermal degradation under high strain conditions in range of 200K-1000K. 


As we can in Fig.\ref{fig:ss-all-temp}, stress-strain curve calculated at 200K and 300K are in accordance with recent calculations, obtained  from DFT and MD methods\cite{desousa2021penta}. Our results show roughly two regimes: A linear-elastic regime up 10\% of strain and subsequently, a plastic region 
resulting from permanent deformations due to structural bond breaking followed by fracture where stress abruptly drops to zero. Similar trends where obtained
at 400K and 500K where there is a progressive reduction of critical fracture points (Ultimate tensile stress $\sigma_{US}$ and critical strain $\varepsilon_c$) One should note that increase in temperature does not significantly influence the values of Young's modulus, but it does affects drastically the critical fracture points in agreement with the kinetic theory of solid fracture \cite{zhurkov1965kinetic,xiao2004kinetic,pei2014effects,li2019effects}. Those values are compiled in Table \tablename{ \ref{tab:elastic-val}}. These critical points decrease with increasing temperature. The Young's modulus obtained in a linear regime ($5$ \% deformation) change in the range of $ 154 - 227 GPa.nm $ for temperatures up to $1000K$. Values found for Young's modulus at 200K-700K (where PG structure remains almost unchanged) are close to values found in the literature by using MD calculations and also according to DFT calculations\cite{desousa2021penta}.

\begin{table}[htb!]
\caption{Elastic values of the PG membrane up to a temperature of 1000K. $Y_{Mod}$ was obtained for a linear regime of 5\%.}
\centering
\begin{tabular}{|c|c|c|c|}
\hline
\textbf{Temperature $(K)$} & $\boldsymbol{Y_{Mod}}$(GPa.nm) & $\boldsymbol{\epsilon_c}$ & 
$\boldsymbol{\sigma_{US}}$ (GPa.nm)\\ 
\hline
200K  &  $222.80 \pm1.95$ &  0.22   & $35.88 \pm0.19$\\ \hline
300K  &  $218.97 \pm2.56$ &  0.20   & $32.58 \pm0.22$\\ \hline
400K  &  $223.84 \pm2.64$ &  0.18   & $31.31 \pm0.26$\\ \hline
500K  &  $224.82 \pm3.07$ &  0.17   & $30.01 \pm0.21$\\ \hline
600K  &  $227.15 \pm3.17$ &  0.13   & $26.94 \pm0.37$\\ \hline
700K  &  $218.16 \pm3.20$ &  0.10   & $23.00 \pm0.29$\\ \hline
800K  &  $224.40 \pm3.33$ &  0.08   & $17.62 \pm0.43$\\ \hline
900K  &  $205.88 \pm5.60$ &  0.06   & $13.26 \pm0.49$\\ \hline
1000K  &  $154.76 \pm3.81$ &  0.06   & $11.83 \pm0.51$\\\hline
\end{tabular}
\label{tab:elastic-val}
\end{table}

\begin{figure}[htb!]
 \centering
 \includegraphics[scale=0.38]{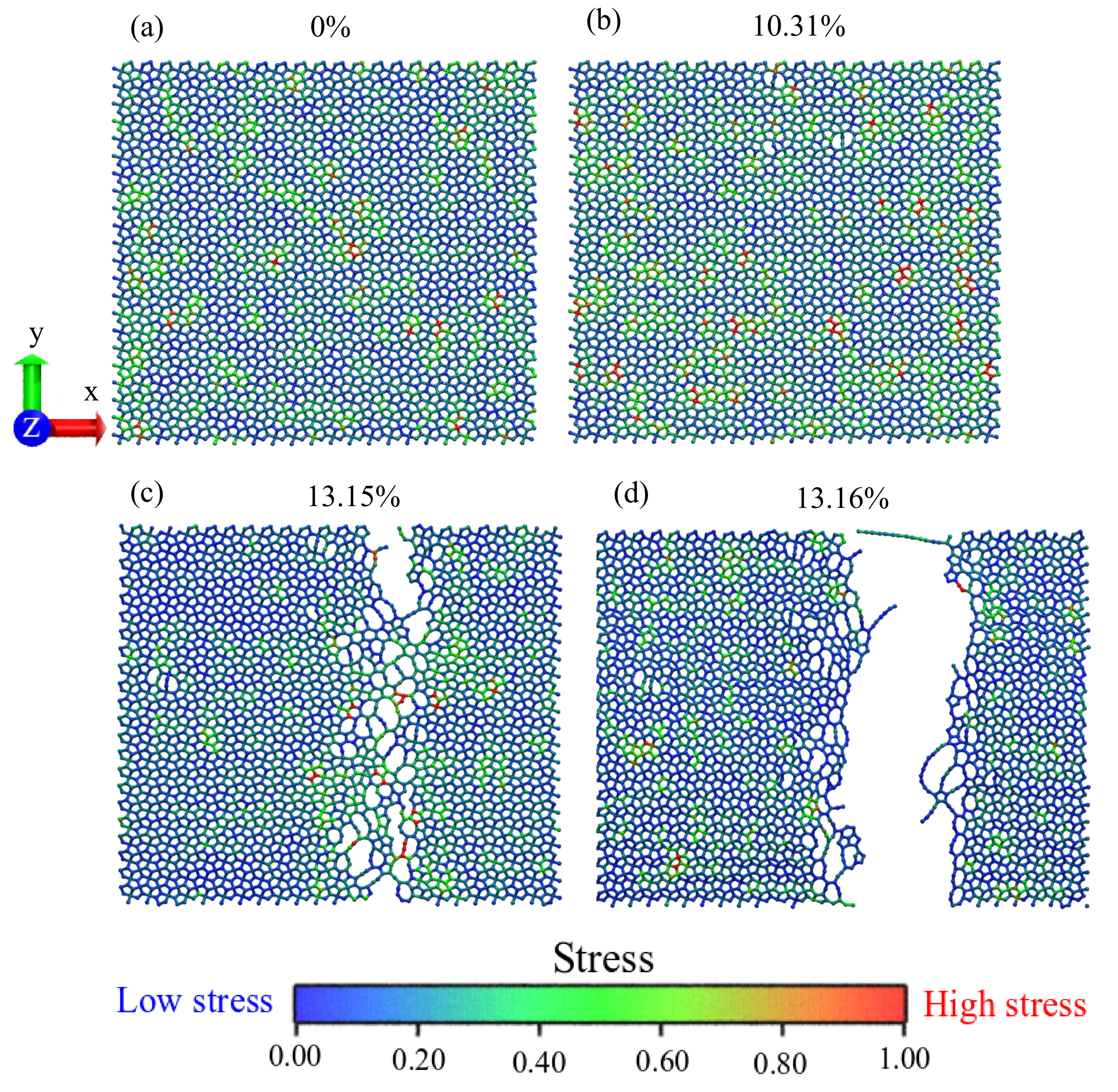}
 \caption{Atomistic representative frames snapshots of reactive molecular dynamics simulations of fracture of the tensile stretch of penta-graphene monolayer at room temperature $600K$. In (a) the penta-graphene monolayer at null stress $0$\% of strain. In (b) strained at $10.31$\%, (c) the crack propagation of nanofracture of penta-graphene monolayer at $13.15$\%. (d) Complete fracture at $13.16$\% of strain. The horizontal color bar represents the level of von Mises stress, where the color blue (red) represents low (high) stress.}
\label{fig:effects-600K}
\end{figure}

The structural transition of PG membrane between 600K and 700K shows a difference in the fracture pattern, confirming the beginning of structural disordering as discussed previously (see Fig. \ref{fig:effects-temp} (c)-(d)). For temperature above $700K$, the penta-graphene monolayer loss of atomic configurations during mechanical stress
and it present only a linear regime followed now by a plastic regime with a smooth stress drop due to the structural reordering at high temperatures. 
The details of fracture patterns of penta-graphene monolayer under stretching and thermal effects at 600K can be visualized in Fig. \ref{fig:effects-600K}, where 
we show the PG nanostructure at four stages of stretch dynamics from null strain (0)\% up to complete fracture of monolayer. The fracture strength dynamics is analyzed through of the  von Mises Stress throughout the penta-graphene monolayer, whith color scheme in In Fig. \ref{fig:effects-600K}. At 10.31 \% of strain (Fig. \ref{fig:effects-600K} (b)) it is possible to see the presence of structural defects and beginning of crack propagation, and the subsequent frames (Fig. \ref{fig:effects-600K} (c) and (d)) illustrates the rearrangement of carbon atoms in the regions from which the crack propagation has originated.

As expected, with the increase of temperature, we observe the fragility increasing of penta-graphene membranes. This effect of temperature can be quantified by the critical strain $\epsilon_{c}$, where for $300K$, $600K$ and $900K$ we obtained values of $20$\%, $13$\% and $6$\%, respectively (see Table \tablename{ \ref{tab:elastic-val}}). Changes due to the temperatures effects on the mechanical properties of PG monolayer were also observed at ultimate stress (US), where for the temperatures above mentioned, we observed a reduction from $\sigma_{US}= 35.88GPa.nm$ at 300K to $\sigma_{US}=26.94GPa.nm$ at 600K and $\sigma_{US}=13.26GPa.nm$ at 900K. Our results show that temperature effects in the fracture dynamics observed for temperatures below $600K$ is characterized by abrupt structural failure followed by sudden fast crack propagation, which is not present for higher temperatures. So, our results distinct thermal effects below and above 600K on the mechanical properties of penta-graphene monolayers. 


Finally, we show in Fig. \ref{fig:effects-1000K} the evolution of fracture of the PG membrane at 1000K.
As discussed in previous section, for temperatures after 900K, after long-time thermalization, the PG membrane experiences a strong structural transition to a disorded membrane. However, during the beginning of stretching dynamics, at 0\% of strain (Fig. \ref{fig:effects-1000K}(a)), we see residual tensions due to the non-equilibrium dynamics. With the propagation of tension over the PG surface, the several bond breaks and pores are observed at 14.18\%(Fig. \ref{fig:effects-1000K}(b)). The C-C bond breaking and pores are distributed along the structure at 27.74\% (Fig. \ref{fig:effects-1000K}(c)). However, no cracking of membrane are observed until an complete fracture in 74.95\% (Fig. \ref{fig:effects-1000K}(e)), where linear carbon chains are formed. The zoomed parts in Figs. \ref{fig:effects-1000K}(g) and (h) show the bond length alternation of C-C bonds which is known to be present carbon chains in its polyyne conformation.



\newpage
\section{Conclusions}

We have studied the effects of temperature in fracture toughness and degradation of Penta-graphene monolayer under uniaxial tensile deformation in $x$ direction  carried out fully atomistic molecular dynamics simulations with interatomic potential ReaxFF implemented in LAMMPS code. Our results show that penta-graphene monolayer strain rate was strong effect on the mechanical properties with reduction of the $67$\%, reduction  in the Ultimate Tensile Strength of the $ 35.88 - 11.83GPa.nm $ and Young's Modulus of the $ 227 - 154GPa.nm $ due to effects temperature. We further investigated the effect of temperature in penta-graphene monolayer at temperatures degradations ranges of $10K$ up to $2000K$ whith loss atomic configurations with temperature effect up to $600K$, where the monolayer show nanostructural transition with islands of graphene, porosity, chains, negative curved layer. We hope that through this work, the nanostructural and mechanical properties of monolayer penta-graphene will be better understood in order to expand their applicability in nanotechnology.

\section*{Acknowledgements}

This work was supported in part by the Brazilian Agencies CAPES, CNPq
and FAPESP. J.M.S acknowledges CENAPAD-SP (Centro Nacional de Alto Desenpenho em São Paulo - Universidade Estadual de Campinas - UNICAMP ) for computational support process (proj842). A.L.A. acknowledges CNPq (Process No. $427175/20160$) for financial support. W.H.S.B., J.M.S and A.L.A. and  thank the Laboratório de Simulação Computacional Cajuína (LSCC) at Universidade Federal do Piauí for computational support.
J.M.S. would like to thank Prof. Douglas Soares Galvão (Full Professor at the Applied Physics Departament - University of Campinas - UNICAMP), for years mentoring my Ph.D., teaching the fundamental basis of Computational Materials Science, kind support as well as being himself an example of ethical and professional conduct that I will keep for the rest of my professional life.
\newpage
\bibliography{penta.bib}

\end{document}